\documentclass[aps,pre,twocolumn,groupedaddress]{revtex4-1}

\usepackage{amssymb}
\usepackage{amsmath}
\usepackage{graphicx}
\usepackage{dcolumn}
\usepackage{bm}
\usepackage[english]{babel}
\usepackage{color}
\usepackage{enumitem}
\usepackage{soul}
\usepackage{epstopdf}

\newcommand{\td}{\mathrm{d}}
\newcommand{\xx}{\boldsymbol{x}}
\newcommand{\vv}{\boldsymbol{v}}
\newcommand{\FF}{\boldsymbol{F}}

\begin{document}

\title{Diffusion in the presence of a local attracting factor:\\
Theory and some interdisciplinary applications}

\author{Hardi Veerm\"ae}
\affiliation{NICPB-National Institute of Chemical Physics and Biophysics, R\"avala 10, 10143 Tallinn, Estonia}

\author{Marco Patriarca}
\affiliation{NICPB-National Institute of Chemical Physics and Biophysics, R\"avala 10, 10143 Tallinn, Estonia}

\date{\today}

\begin{abstract}
\noindent
We study a simple model of a random walker in $d$ dimensions  moving in the presence of a local heterogeneous attracting factor expressed in terms of an assigned space-dependent ``attractiveness function'', a situation frequently encountered in the study of various diffusion problems.
The corresponding drift-diffusion equation and the explicit expressions for the velocity field and the diffusion coefficient are obtained and discussed.
We consider some examples of applications of the results obtained to chemotactic diffusion processes and social dynamics.
\end{abstract}


\maketitle

\section{Introduction}
The theory of Brownian motion and random walks represents a cornerstone of statistical mechanics for providing an underlying physical picture and a microscopic explanation of many diffusion processes. The most famous examples are probably those provided by the theory of Brownian motion of Einstein~\cite{Einstein1905a,Einstein1906b,Einstein_1956_A} and Smoluchowski~\cite{Smoluchowski1906a,Smoluchowski1916a}.
Since then the application range of random walk theory has widened in many different fields, extending to e.g. biology, ecology, and anthropology. 
The standard drift-diffusion equation,
\begin{equation} 
  \label{eq-0}
  \partial_t n(\xx,t)  = 
      \nabla 
      \left[ 
        - n(\xx,t) \vv(\xx,t) 
        + D \nabla n(\xx,t)  
      \right],
\end{equation}
where $\xx = (x_1,x_2,x_3)$ is the particle position vector, defines the dynamics of the Brownian particle through the drift velocity field $\vv(\xx,t)$ and the diffusion coefficient $D$.
In principle $\vv(\xx,t)$ and $D$ can be derived from the knowledge of the microscopic dynamics of the underlying random motion. 
In the well known problem of the overdamped Brownian motion the velocity field turns out to be given by $\vv(\xx,t) = \eta^{-1} \FF(\xx,t)$, where $\eta$ is the friction coefficient, $\FF(\xx,t) = - \nabla U(\xx,t)$ the external force acting on the Brownian particle, and $U(\xx,t)$ the corresponding external potential. The diffusion coefficient $D$ and the friction coefficient $\eta$ are related to each other and to the temperature $T$ of the  environment through the fluctuation-dissipation theorem, $\eta D = k_\mathrm{B}T$~\cite{VanKampen1992a,Gardiner1985a}.

Besides the mentioned works by Einstein~\cite{Einstein1905a,Einstein1906b,Einstein_1956_A} and Smoluchowski~\cite{Smoluchowski1906a,Smoluchowski1916a} various derivations of the drift-diffusion equation have been proposed, starting from a microscopic picture of the corresponding random walk: the interpretation  of the parameters in the resulting drift-diffusion equation depends on the specific problem under study --- see also Van Kampen's derivation of the diffusion equation from a general discrete one-step processes~\cite{VanKampen1992a-VIII}.
In fact, since their very beginning diffusion problems have been appearing  in very different disciplines, as exemplified by the first random walk model introduced in the social sciences by Louis Bachelier in 1900 with the goal of modeling asset price dynamics~\cite{Bachelier_2006_A}.

In many diffusion problems the underlying dynamics of the random walkers is intrinsically different from that of a Brownian particle. In place of an actual drift force, one  often finds an ``attractiveness function'' $\kappa(\xx,t)$ that modulates the otherwise isotropic and homogeneous probability that a random walker jumps at time $t$ from its current position to the generic position $\xx$.
An example is chemotactic diffusion where $\kappa(\xx,t)$ represents the concentration of some substance affecting the drift motion of the random walker, that can be e.g. a cell, a bacterium or an ant.
Even if the attractiveness function $\kappa(\xx,t)$ is not per se a potential, it is shown below that its effect is to generate a term analogous to an external force in the drift-diffusion equation.

Diffusion processes taking places in such situations, in which the drift motion can be traced back to some attracting factor rather than to a (known) force, represent the subject of the present paper.
The basic difference is that in the latter case the velocity field, e.g. the potential gradient in the case of Brownian motion, provides by definition the actual flux of particles or the average velocity of the Brownian particle.
Instead, the attractiveness function $\kappa(\xx,t)$ embodies the effects of some external factor on the preferences of the random walker and therefore \emph{vanishes in the absence of diffusion}.
In the type of problems considered here, the random walker \emph{compares} the (known) values of some attractiveness measure $\kappa(\xx,t)$ in the accessible positions before making the next jump, choosing the jump direction in a probabilistic way depending on the result of the comparison made.

The goal of the present paper is to study the general drift-diffusion equation for such type of systems directly in terms of the attractiveness $\kappa(\xx,t)$. 
To this aim we proceed through two steps: 

We start from a general random walk in a $d$-dimensional space where jumps are defined by a jump probability $p_0(x,t|x',t')$ from position $x'$ at time $t'$ to position $x$ at time $t$. The mathematical preliminaries are laid out in Sec.~\ref{pre}.
Then we introduce an attractiveness function $\kappa(x,t)$, assuming that the probability that the jump takes place along a certain direction is proportional to the local value of the attractiveness function.
The corresponding drift-diffusion equation is derived in Sec.~\ref{attraction}.
It is shown that to this aim one can simply replace the conditional jump probability  $p_0(x,t|x',t')$ with the corresponding \emph{conditional expectation}  $p(x,t|x',t')$ obtained by weighting $p_0(x,t|x',t')$ with the attracting function $\kappa(x,t)$ --- see Sec.~\ref{derivation}.
The probabilistic meaning of the attracting function $\kappa(x,t)$ is further discussed in Sec.~\ref{stat}.

The motivation for the present investigation is that the theoretical framework outlined above and the corresponding diffusion equation obtained well suit the description of many diffusion processes in which the appearance of a drift motion can be traced back to a non-constant function $\kappa(x,t)$ measuring a local preference or attractiveness.
The equation obtained is in fact simple, yet general enough, to be useful for applications to diffusion processes in physics and in interdisciplinary fields.
Some applications to chemotaxis and social dynamics are  considered in Sec.~\ref{applications}.
Possible developments and further applications are discussed in section~\ref{conclusions}.

\section{Derivation of the diffusion equation}
\label{derivation}

It is possible to give a general derivation of the drift-diffusion equation from a random walk model starting from an arbitrary form of the joint step-length and waiting-time distribution of the random walker~\cite{Othmer1988a,Klafter1987a}.
Here we adopt a similar procedure by first considering a generic jumping probability $p_{0}(x,t + \tau|x',t)$ that describes the motion of the random walker in the absence of the attracting factor. Thereafter the effect of the attracting factor on this random motion is studied and the corresponding drift-diffusion equation is derived with the Kramers-Moyal procedure.

\subsection{Mathematical preliminaries}
\label{pre}

Here and in the following we consider the general case of a $d$-dimensional space, in which each point $x$ is specified in terms of its $d$ coordinates, $x = (x_1, x_2, \dots, x_d)$. If $p(x,t + \tau|x',t)$ is the probability that a random walker at location $x'$ at time $t$ hops to location $x$ where it will be located at time $t+\tau$, the random walker density $n(x,t+\tau)$ at $t+\tau$
can be obtained from the density $n(x',t)$ at time $t$ as
\begin{align}\label{n1}
   n(x, t + \tau) 
   & = \int \! \td^{d}x' \, p(x,t + \tau|x',t) \, n(x',t) \, . 
\end{align}
The normalization condition
\begin{equation}
\int \td^{d}x \, p({x},t|{x}',t') = 1
\label{pnorm}
\end{equation}
is assumed to hold for any $x'$ and $t > t'$.

To have a simple well behaved continuum limit, where the dynamics can be approximated by a diffusion equation, we impose the condition of \emph{locality}: there exists a characteristic length scale $\epsilon$, so that jumps with distances exceeding this scale are very improbable.  Formally we require for the probability of long jumps to be at least exponentially suppressed:
\begin{align}\label{eq:locality_cond}
	p_{0}(x + y|x) < C \exp(-|y|/\epsilon), \quad \mbox{if} \quad |y| > \epsilon.
\end{align}
Time arguments will be omitted in the following, i.e. $p(x|x') \equiv p(x,t + \tau|x',t)$. It is shown in appendix \ref{app:KM_bound}, that this form of locality defined by Eq.~(\ref{eq:locality_cond}) is sufficient for the Kramers-Moyal coefficients $D_{i_1\ldots i_m}$ to be at least of order $\epsilon^{m}$. From the Kramers-Moyal procedure we obtain that the behavior of the random walkers at large scales is described by a drift-diffusion like equation of the form (see App. \ref{app:KME_derivation})
\begin{align}\label{DD0}
	\partial_t n 
	= \partial_{i} [ D^{(0)}_{ij} \partial_{j} n - D^{(0)}_{i} n ]
	+	\mathcal{O}\left(\epsilon^{3}/\tau\right).
\end{align}
Thus  in the continuum limit, i.e. when $\epsilon \to 0$ with $\epsilon^{2}/\tau$ fixed, the higher order terms vanish and the diffusion equation is obtained in its canonical form. Above $\partial_{t} \equiv \partial/\partial t$, $\partial_{i} \equiv \partial/\partial x_{i}$ and summation over repeated indices is assumed. $D^{(0)}_{i}$, $D^{(0)}_{ij}$ are the drift and the diffusion terms corresponding to $p_0(x|x')$:
\begin{align}
	D^{(0)}_{ij} &= \lim_{\epsilon \to 0}\frac{1}{2\tau}\int \td^{d} y \, p_{0}(x + y|x) y_{i}y_{j},
	\\
	D^{(0)}_{i} &= \lim_{\epsilon \to 0}\frac{1}{\tau}\int \td^{d} y \, p_{0}(x + y|x) y_{i} - \partial_j D^{(0)}_{ij}
\end{align}

We stress that the characteristic length $\epsilon$ of the steps and a fixed time interval $\tau$ between the steps are the only restrictions imposed the motion of the random walker. Thus our set-up is fairly general as, at the macroscopic level, it can be applied to any drift-diffusion equation.

\subsection{Drift from an attracting factor}
\label{attraction}

In the following we assume the existence of an external inhomogeneous effect, due to the local landscape attractiveness, on top of the random motion assumed above and study its effects on the drift and diffusion of the Brownian particle.
The attracting factor is represented by the function $\kappa(x,t)$, that is assumed to modulate the diffusion so that the transition probability becomes proportional to it, i.e.,
\begin{align} \label{p_k}
p(x|x') \propto \kappa(x,t) \, .
\end{align}
Since $\kappa(x,t)$ is not a probability but only modulates it, it has to be normalized  in order to obtain the jump probability: the normalization conditions require that the driftless hopping probability $p_0(x|y)$ is replaced by the conditional expectation value~\cite{Feller1966a} weighted by the attracting function $\kappa(x,t)$,
\begin{align} \label{p1}
	p_0(x|x') \to
        p(x|x') = \frac{\kappa(x)p_0(x|x')}{\int \td^{d} x'' \kappa(x'') p_0(x''|x')}.
\end{align}
This new hopping probability $p(x|x')$ still obeys the normalization condition of Eq.~(\ref{pnorm}) and reduces to the unperturbed form $p_0(x|x')$ when $\kappa$ is a constant.
We stress that the normalization factor in the denominator of Eq.~(\ref{p1}) is a $\kappa$-weighted sum of the hopping probabilities $p_0(x''|x')$ over all the possible \emph{arrival} positions $x''$, starting from the same position $x'$, see Fig.~(\ref{fig_2D}).
In a typical situation the quantity $\kappa(x,t)$ can be directly related to some observable quantities, for example to pheromone density in the case of chemotactic diffusion discussed below.

A drift term is now induced by the attracting factor $\kappa(x)$ since according to Eq.~(\ref{p1}) $p(x|y) \propto \kappa(x)$.
Following the Kramers-Moyal procedure (see App.~\ref{app:KME_derivation}), the expansion of $\kappa$ around $x$ yields
\begin{align}
\label{Dij}
	D_{ij}(x) 
&	=	\frac{1}{2\tau}
		\frac{\int \td^{d} y \, y_{i}y_{j}p_{0}(x+y|x)\kappa(x+y)}{\int \td^{d} y \, p_{0}(x + y|x)\kappa(x+y)}
	\nonumber\\
&	=	D^{(0)}_{ij}(x) + \mathcal{O}(\epsilon),
	\\
\label{Di}
	D_{i}(x) 
&	= 	\frac{1}{\tau}
		\frac{\int \td^{d} y \, y_{i} p_{0}(x+y|x)\kappa(x+y)}{\int \td^{d} y \, p_{0}(x + y|x)\kappa(x+y)} - \partial_j D_{ij}
\nonumber\\
& 	=	D^{(0)}_{i}  + 2 \kappa^{-1} D_{ij}\partial_{j}\kappa + \mathcal{O}(\epsilon)
\end{align}
%
Thus, in the hypothesis that $n(x,t)$ and $\kappa(x,t)$ vary slowly in space (at scales comparable to $\epsilon$) and neglecting higher order terms in $\epsilon$ (taking the continuum limit), the Kramers-Moyal procedure yields an additional drift term
\begin{align}\label{Dkappa}
	D^{(\kappa)}_{i}(x) 
&	\equiv	2 D_{ij}(x)\partial_{j}\ln(\kappa(x))
\end{align}
corresponding to the effective modulated transition probability, while negligibly the modifying the diffusion term. In conclusion, the following drift-diffusion equation is obtained:
\begin{align}\label{DD1}
	\partial_t n = \partial_{i} [ D_{ij} (\partial_{j} n - \partial_{j}\ln(\kappa^{2}) n) - D^{(0)}_{i} n ] .
\end{align}
The drift-diffusion equation \eqref{DD1} can also be recast as
\begin{align}\label{DD2}
	\partial_{t} n 
	=	\partial_{i} \left[\kappa^2 D_{ij} \partial_{i}(\kappa^{-2}n) - D^{(0)}_{i} n\right] .
\end{align}
It follows, that the absence of the background drift term, i.e. when $D^{(0)}_{i} = 0$, the equilibrium distribution is given by $n_{eq} \propto \kappa^{2}$ if we assume a time independent $\kappa$. However, in many practical applications $\kappa$ itself can be a dynamical quantity and thus the assumption of time independence might not hold at all or it could hold only approximately, e.g. when $\kappa$ changes slowly enough for the density $n$ to remain close to the equilibrium.

Finally we note that the drift term in \eqref{Dkappa} is proportional to the the  diffusion tensor $D_{ij}$, thus rendering drift caused by an attracting factor $\kappa$ conceptually different from a drift attributed to an external force. Especially, in the absence of diffusion, i.e. when $D_{ij} = 0$, the drift will vanish too ---  in this respect see also section \ref{stat}.

\subsection{A example: free diffusion with an attracting factor}
\label{example}

\begin{figure}
\includegraphics[width=7cm]{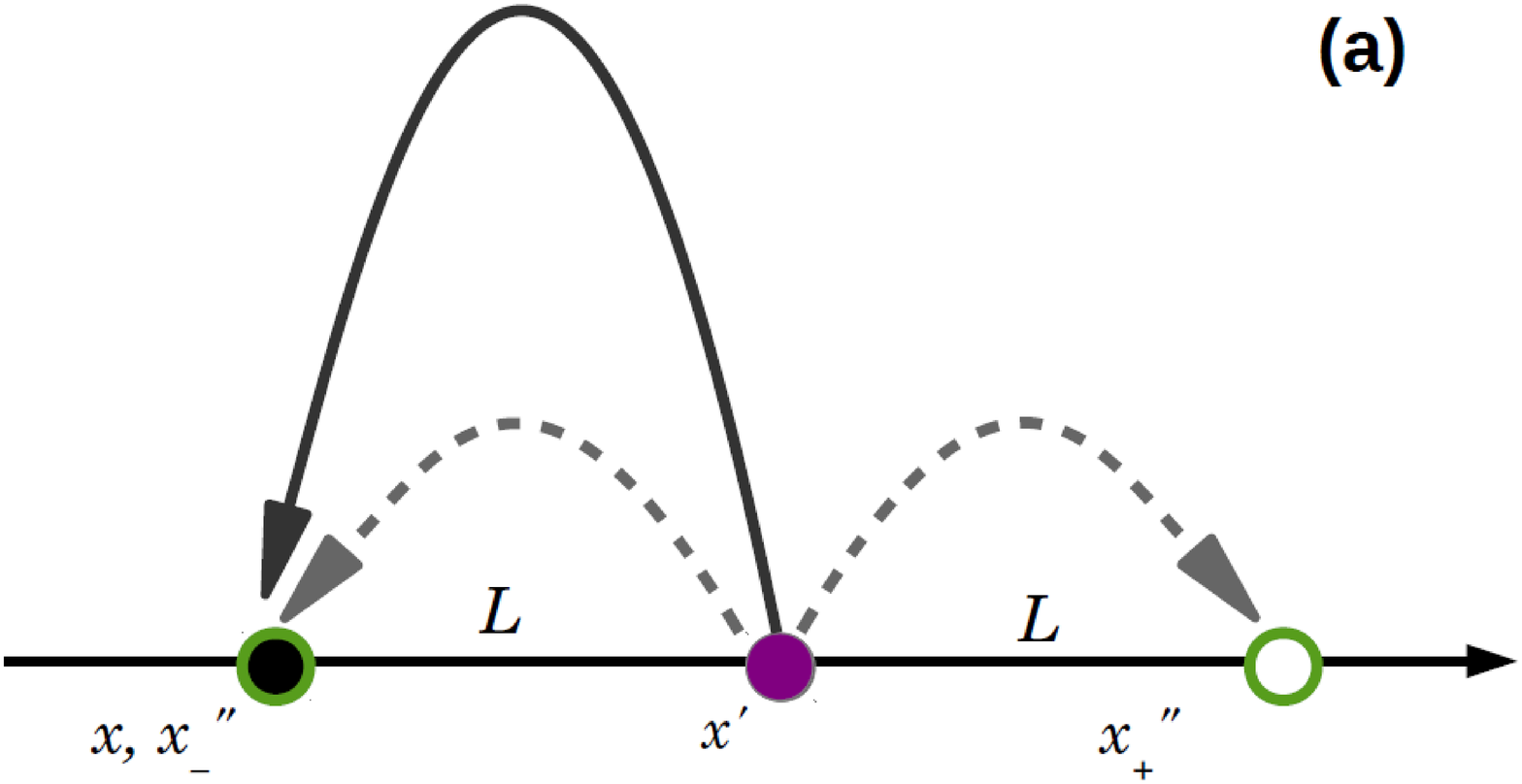}
\includegraphics[width=7cm]{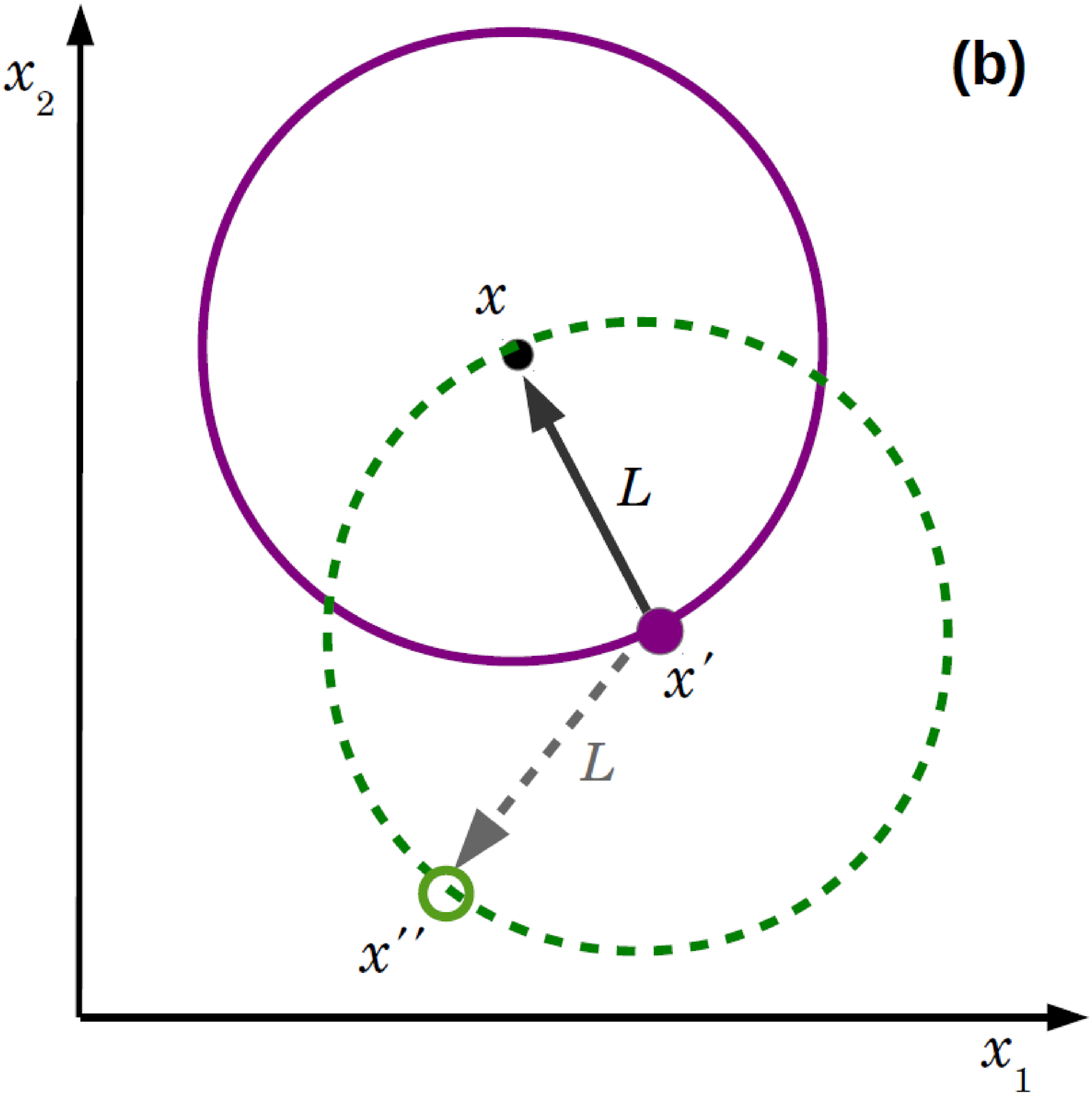}
\caption{\label{fig_2D}
Scheme of a diffusion process with constant step $L$ in (a) one and (b) two dimensions (Rayleigh problem).
Before jumping from the initial position $x'$ to the arrival position $x$, the random walker probes the environment, i.e. the values of the function $\kappa(x)$ in all the neighboring locations $x''$ at a distance $L$ from $x'$.
The resulting probability $p(x|x')$ to jump from $x'$ to $x$ is the conditional expectation~(\ref{p1}).
}
\end{figure}

For illustrative purposes let us consider the common example, where, in the absence of an attracting factor, the random walker would follow a homogeneous and isotropic transition probability of the form
\begin{align}\label{eq:free_dif_p}
	p(x,t+\tau|x',t) \propto \delta(|x'-x| - \ell),
\end{align}
In this case the motion of the random walker consists of consecutive steps of constant length $\ell$ made at regular time intervals $\tau$ but with no preferred direction. In two dimensions this coincides with the Rayleigh problem that is illustrated in Fig.~\ref{fig_2D}. The diffusion constant related to (\ref{eq:free_dif_p}) reads
\begin{align}
	D_{ij} = \frac{\ell^{2}}{2\tau d} \delta_{ij}
\end{align}
where $d$ denotes the dimension of the underlying space. In all, in the absence of an attracting factor, the random walker is described by the most common form of the diffusion equation, $\partial_{t} n = D \nabla^{2} n$.
 
In the presence of an attracting factor, the random walker can choose to jump to any point on the circle in Fig.~\ref{fig_2D}, but before making a choice he probes the environment and adjusts his choice so that locations with an higher value of the attracting factor are preferred -- as a result drift is created. However, the motion of the random walker is still driven by diffusion that remains unmodified by the attracting factor. The resulting drift-diffusion equation can be obtained from Eq.~\eqref{DD1}. It reads
\begin{align}\label{DD3}
	\partial_{t} n 
	=	D \nabla(\nabla n - n\nabla\ln(\kappa^{2})).
\end{align}
This equation is formally equivalent to the drift-diffusion equation~(\ref{eq-0}) with the velocity field given by $v_{i}(x,t) = 2 D \partial_{i} \ln\kappa(x,t)$, corresponding to an effective external potential 
\begin{align}
	U(x,t) \propto -\ln\kappa(x,t).
\end{align}
The last relation could also be obtained from the equilibrium conditions where $n_{eq}(x) \propto \exp(-\beta U(x))$, but on the other hand $n_{eq}(x) \propto \kappa(x)^{2}$.

\subsection{Statistical Interpretation}
\label{stat}
%
At a general level, Eq.~(\ref{p1}) describes a modulated jump probability that can have different origins.
To have a concrete example in mind we can think of some ecological dispersal problem in which an individual checks e.g. the fitness of the local landscape before taking the next move.
In other problems of chemotactic diffusion, the cell or a bacterium that represents the random walker must be able somehow to monitor the surrounding environment in order to get information about the concentration of some chemical substance.
In all these cases there is a common relevant physical principle underlying the various interpretations of the conditional probability~(\ref{p1}): namely, the elementary jump process described by Eq.~(\ref{p1}) can  be given a probabilistic interpretation in which a random walker first \emph{compares} some property of the neighboring zones and thereafter chooses the direction of the next step (randomly but with probabilities based on the comparison made).

The operative meaning of the conditional expectation~(\ref{p1}) has been studied by Othmer and Stevens~\cite{Othmer-Stevens-1997a}.
Their comparison of various 1D and 2D lattice models of chemotactic diffusion clearly shows that the normalization implied by the conditional expectation~(\ref{p1}) is the key-element for obtaining a Weber-Fechner-like logarithmic dependence of the chemotactic potential --- analogous to the potential  $U(x)$ of Brownian motion --- on the relevant chemical concentration, corresponding here to the attractiveness function $\kappa(x)$.

Notice that even in those cases in which the random walker performs a comparison based on the ratio $\kappa(x)/\kappa(x')$, expressing the attractiveness of the final position $x$ \emph{relative} to that of the starting location $x'$ --- rather than on the arrival values $\kappa(x)$  alone --- one still obtains the same expression as in Eq.~\eqref{p1}:
\begin{align} \label{p2}
p(x|x') & = \frac{p_0(x|x')\kappa(x)/\kappa({x}')}{\int \td^{d} x'' p_0(x''|x')\kappa(x'')/\kappa(x')}
\nonumber\\
        & \equiv \frac{p_0(x|x')\kappa(x)}{\int \td^{d} x'' p_0(x''|x')\kappa(x'')} \, ,
\end{align}
due to the cancellation of the $\kappa(x')$ factors.

In the typical situations considered in this paper the drift-diffusion equation, Eq.~(\ref{DD1}), is obtained from the microscopic random walk dynamics with a given attractiveness function $\kappa(x)$.
Before proceeding, it is useful to make a simple consistency check of the results obtained thus far by considering the complementary situation, i.e., starting from the drift-diffusion equation (\ref{eq-0}) for an overdamped Brownian particle and asking what is the form of the corresponding attractiveness function $\kappa(x)$.
Comparison with Eq.~(\ref{DD1}) shows that the external potential felt by the random walker is proportional to the logarithm of the attractiveness, i.e.,  $U(x,t) = - \alpha \ln \kappa(x)$, where $\alpha > 0$ is some constant.
The definition of the attracting function $\kappa$ implies that the transition probability from a starting point $x'$ to an arrival point $x$ is proportional to $\kappa(x)/\kappa(x') = \exp\{ - [U(x) - U(x')]/\alpha \}$, which coincides with the Boltzmann-Gibbs factor (with $\alpha$ representing temperature), used e.g. in the Monte Carlo method~\cite{FRENKEL1996a_MC}.

The same procedure can be useful when one starts from an assigned drift-diffusion equation and searches a corresponding microscopic random walk model defined in terms of an attracting function $\kappa(x)$.

\section{Applications}
\label{applications}

\subsection{Application to chemotaxis (Keller--Segel model)}
\label{chemotaxis}
%
An example of non-isotropic diffusion is that driven by chemotaxis, a mechanism determining many types of biological motion based on the concentration of a certain chemical.
The motion of units is usually biased toward higher values of the  concentration of a certain chemical,
referred to as pheromone.
Many instances of this kind of motion are known in different forms of life, with units represented by animals, insects, bacteria, and cells and the pheromone by the substance they or other sources produce; 
for example, leukocyte cells in the blood move towards a bacterial inflammation following the positive gradient of some chemical caused by the infection;
slime mold {\it Dictyostelium Discoideum}, a single-cell amoebae, moves towards regions  of high concentrations of the cyclic-AMP produced by the amoebae themselves~\cite{Murray2003a_11}.

A standard and simple model of chemotactic dynamics was put forward by Keller and Segel~\cite{Keller1970a,Murray2003a_11},
\begin{subequations}
\begin{eqnarray}
\label{KS1}
  \frac{dn}{dt} 
    &=&
    f(n) 
    + \nabla \cdot [D \nabla n] 
    - \nabla [n \chi(a) \nabla a] ,
    \\
\label{KS2}
  \frac{da}{dt} &=&
    g(a,n) 
    + \nabla \cdot [D_a \nabla a] 
    \, .
\end{eqnarray}
\end{subequations}
Here $n(x,t)$ is the density of diffusing biological units and $a(x,t)$ is a dimensionless quantity proportional to the concentration of pheromone.
The function $f(n)$ in Eq.~(\ref{KS1}) defines the type of population dynamics characterizing the units and $g(a,n)$  in Eq.~(\ref{KS2}) the production of pheromone possibly due to the units themselves, as in the mentioned example of the {\it Dictyostelium Discoideum}, as well as its decay.
The second terms on the {\it rhs} of Eqs.~(\ref{KS1}) and (\ref{KS2}) describe the diffusion process undergone by the units and the pheromone, with diffusion coefficients $D$ and $D_a$ respectively.
Finally, one has to assign the form of the last term on the {\it rhs} of Eq.~(\ref{KS1}) responsible for the drift motion of the units due to the chemotactic substance, i.e. $ \chi(a)$, usually defined as the ``chemotactic coefficient'' or ``chemotactic sensitivity''.
Interpreting the drift term as that of a Brownian particle acted upon by the effective force $\mathbf{F}(x) = \chi(a(x))\nabla a(x)$, one can set
\begin{eqnarray}
   \label{Xa}
   \chi(a) = - \frac{dX(a)}{da} ,
\end{eqnarray}
and rewrite the effective force simply as
\begin{eqnarray}
   \label{Feff}
   \mathbf{F}(x) = \chi(a(x))\nabla a(x) \equiv \nabla X(a).
\end{eqnarray}
Then Eq.~(\ref{KS1}) becomes the drift-diffusion equation of an overdamped Brownian particle in an external potential $U(x,t) = X(a(x,t))$.
Following this analogy we will refer to the quantity $X(a)$ as the ``chemotactic potential''.
By comparison between Eqs.~(\ref{KS1}) and (\ref{DD1}) and using Eq.~(\ref{Xa}) one can see that the attractiveness function is in this case given by
\begin{eqnarray}
   \label{K-X}
   \kappa(x,t) = \kappa_0 \exp\left( - \frac{X(a(x,t))}{2D}\right) \, ,
\end{eqnarray}
where $\kappa_0$ is some constant.
According to this picture, units diffuse in space with a diffusion probability modulation represented by Eq.~(\ref{K-X}).

In their original paper, Keller and Segel showed that in order for Eqs.(\ref{KS1}) and (\ref{KS2}) to reproduce the waves observed experimentally in microbe diffusion processes the chemotactic coefficient must have a singularity of order one or larger, that is, if one chooses a simple power law form $\chi(a)  = 1/(a + a_0)^b$, where $a_0$ is a constant and $b \ge 1$. 
Keller and Segel chose $a_{0} = 0$ and the least singular of all the possible powers, $b = 1$, corresponding to a logarithmic chemotactic potential,
\begin{eqnarray}\label{a-chi}
  X(a) = - \ln (a) \, .
\end{eqnarray}
In this case Eq.~(\ref{K-X}) shows that the attractiveness function is simply a power of the pheromone concentration.
In fact such a dependence is found e.g. in E.Coli at least in a relevant part of the parameter ranges~\cite{Kalinin2009a}. 


Different laws and forms of the drift-diffusion equation can be obtained by choosing correspondingly different forms of the attractiveness function. In some studies chemotactic potentials different from the one with a minimal exponent value $b=1$~\cite{Kalinin2009a}, corresponding to the Weber-Fechner (or log) law, have been proposed. The value $b \approx 2$ can be justified on the base of experimental evidence~\cite{Lapidus1976a} ---  moreover the applicability of the Weber-Fechner law seems to depend on the parameter range of the system under study. The case $b = 2$ corresponds to a chemotactic coefficient and potential
\begin{eqnarray}
   \label{b2}
   \chi(a) = \frac{1}{(a + a_0)^{2}} \, ,
    ~~~~
    X(a) = \frac{1}{a + a_0} \, ,
\end{eqnarray}
referred to as the receptor law. 
Besides the decrease of the effect of pheromone concentration at very low concentrations $a \ll a_0$, this form of $X(a)$ avoids the infinite response that would be otherwise obtained in the case of a logarithmic chemotactic potential when $b = 1$ for $a \to \infty$, since, on the other hand, here $X(a \!\to\! \infty) \to 0$. The chemotactic potential (\ref{b2}) is associated to the attractiveness function $\kappa(x,t) = \kappa_0 \exp\{ - 1 / [2D(a_0 + a(x,t))] \}$ that consistently exhibits the ineffectiveness of pheromone at very low concentrations as well as the saturation effect at high concentrations.

\subsection{Population Dispersal}
\label{dispersal}

The function $\kappa(x,y)$ may also represent the relative difficulty or easiness in following a certain path in a 2D diffusion process across a surface.
Diffusion of ions, small and large molecules on the surface of biomolecules provide relevant examples of such inhomogeneous diffusion. 
In this case the Kramers-Moyal coefficients $D_{ij}$ are directly related to the entropic forces due to the boundaries.

Other examples come from the studies of dispersal of animals and humans across a geographical landscape.
There are many factors related to the physical geography which could be well described in terms of a modulation of diffusivity.
In all these cases there is usually no net systematic drift due to some sort of forces pushing individuals along certain directions, i.e., it is difficult to formulate an analogue of the external potential acting on a Brownian particle.
On the other hand, it is natural to introduce (and possibly measure experimentally) the inhomogeneous probability that a certain path rather than another one is followed and express such a probability in terms of a local attracting factor.
For instance, paths that follow sea and river shores are known to represent corridors characterized by a higher mobility, while the same shores or mountains interrupting a path are locally associated to lower effective diffusivities~\cite{Lomolino2006a}.

In the dispersal of human or animal groups, diffusion may become non-linear, since the presence of other individuals will condition the diffusion process of each single  individual deciding in which direction to make the next diffusive step.
For instance, in the well known example of bacteria diffusion, a non-linear diffusion equation~\cite{Murray2002a} of the form
\begin{equation}\label{NL}
\frac{\partial n}{\partial t} = \nabla\left[ D(n) \nabla n \right] \, ,
\end{equation}
describes an inhomogeneous diffusion depending on the density $n(x,t)$ itself in which individuals escape regions of higher density faster, if $D(n)$ is a growing function of the density $n$.
The diffusion equation \eqref{NL} can be put into the form of Eq.~\eqref{DD2}, 
$$
	\partial_{t} n = D_{0} \nabla(\nabla n - n\nabla\ln(\kappa^{2})),
$$
with
\begin{align}
	\kappa^{2} &= \frac{n}{n_0} \exp\left(-\int \td n\frac{D(n)}{D_0 n}\right).
\end{align}

A standard form of the density-dependent diffusion coefficient is given by $D(n) =  D_0 (n / n_0)^q$, where $D_0$, $n_0$, and $p$ are suitable positive parameters~\cite{Murray2002a}. Interestingly, in this case the corresponding attracting factor,
\begin{align}
	\kappa^{2} 
	= \frac{n}{n_0} \exp\left(-\frac{1}{q}\left(\frac{n}{n_0}\right)^q\right),
\end{align}
resembles a stretched Boltzmann-Gibbs shape and reduces to the canonical form for $q = 1$.

\subsection{Social Dynamics in Space}
\label{innovation}

It is worth mentioning other types of diffusion processes taking place across geographical landscapes, in which the probability that the diffusion process develops along a certain direction can be best related to some local factor rather than to an external force producing a drift.
In this set of problems one can find e.g. processes of epidemic diffusion, opinion dynamics, and (technological) innovation spreading.

Social dynamics of opinion spreading offers a good example.
Many-agents social dynamics models or continuous opinion spreading models such as the model by Borghesi and Bouchaud, introduced to study political elections~\cite{Borghesi2010a}, can provide a detailed description of the system state evolution and the corresponding modulated diffusion process.

The example of the spreading of agriculture during the neolithic age is an interesting example of technological innovation spreading, since it took place across a wide area but under the influence of various factors, such as the same underlying geography affecting human dispersal or the geophysical properties of the land related to its suitability for agricultural purposes --- see Ref.~\cite{Fort2015a} for a recent investigation.

Studies dealing specifically with technological spreading have to take into account the landscape inhomogeneities.
For instance H\"agerstrand's model of innovation diffusion~\cite{Hagerstrand1967a} describes the level of adoption of a new innovation in terms of a mean information field evolving in space and time according to a reaction-diffusion-like equation.
There are various factors making such a process inhomogeneous,  possibly affecting directly the innovation diffusion process, see e.g. Refs.~\cite{Kandler2009a,Shinohara2012a}.

\section{Conclusion}
\label{conclusions}

In this paper we have proposed a framework providing a simple description of a diffusion process influenced by some local factor represented by an assigned attracting function $\kappa(x)$.
This framework may be most suitable to study problems of random motion in which the random walkers are able to compare the possible directions and to use the gathered information to make a (probabilistic) decision about the direction of the next step.
The information about the positions is encoded in the attracting function $\kappa(x)$ so that locations with a higher value of the attracting function are chosen with an higher probability.

We have discussed the example of chemotactic diffusion, also mentioning how the framework proposed easily lends itself to the study of other ecological and dispersal models where the influence of the spatial landscape can be described by a local fitness $\kappa(x,t)$.
For instance, models of dispersal and competition between species, see Ref.~\cite{Tereshko1999a}, or related models where the local density of resource plays a key role~\cite{Tereshko2000a}, are perfect candidates for this type of study.
Also the mathematical modeling of human dispersal, usually done in terms of reaction-diffusion equations (for a recent study see Ref.~\cite{Fort2015a}), rarely takes into account the heterogeneity of the local landscape.
In these and similar situations the drift-diffusion equation framework proposed here may provide a suitable tool for carrying out a more realistic modeling of the actual diffusion process.

\begin{acknowledgments}
This work was supported by the institutional research funding IUT (IUT39-1 and IUT23-6) of the Estonian Ministry of Education and Research.
\end{acknowledgments}

\appendix
\section{A short derivation of the Kramers-Moyal expansion}
\label{app:KME_derivation}

The Fourier transformation of the conditional probability gives the moment generating function (time arguments are omitted):
\begin{align}
	\int \td x' \, \exp(i k (x'-x)) p(x'|x) = \sum^{\infty}_{m=0} \frac{(ik)^{m}}{m!} M_{m}(x),
\end{align}
where $M_{m}(x) := \int \td y \, p(x + y|x) y^{m}$ is the $m$-th moment for jumping from $x$ to $x'$. This corresponds to the following expansion
\begin{align}
	p(x'|x) 
&	= \sum^{\infty}_{m=0} \frac{1}{m!} M_{m}(x)(\partial_{x})^{m}\delta(x'-x)		
\end{align}
Plugging this into the master equation
\begin{align}
	n(x,t + \tau) = \int \td x' p(x,t+\tau|x',t) n(x',t) 
\end{align}
gives the (formally exact) expansion
\begin{align}
&	\frac{1}{\tau}(n(x,t + \tau) - n(x,t))
  \nonumber\\
&	=	\sum^{\infty}_{m=1} (-\partial_{x})^{m}(D_{m}(x,t,\tau) n(x,t)),
\end{align}
where 
\begin{align}
	D_{m}(x,t,\tau) := \frac{1}{m!\tau}\int \td y \, p(x + y,t+\tau|x,t) y^{m}.
\end{align}
In $d$-dimensions:
\begin{align}
&	\frac{1}{\tau}(n(x,t + \tau) - n(x,t))
  \nonumber\\
&	=	\sum^{\infty}_{m=1} (-1)^{m}(\partial_{i_{1}}\cdots \partial_{i_{m}})(D_{i_1\ldots i_m}(x) n(x)),
\end{align}
where 
\begin{align}
&	D_{i_1\ldots i_m}(x,t,\tau) 
	\\
&	:= \frac{1}{m!\tau}\int \td^{d} y \, p(x + y,t+\tau|x,t) y_{i_1}\cdots y_{i_m}
\end{align}

\section{An upper bound on the Kramers-Moyal coefficients}
\label{app:KM_bound}

We prove that the requirement of locality (\ref{eq:locality_cond}) $p(x + y|x) < C \exp(-|y|/\epsilon)$ if $|y| > \epsilon$ guarantees that $D_{i_1\ldots i_m}$ is at least of order $\epsilon^{m}$:
\begin{align}
&	|D_{i_1\ldots i_m}(x)|
\nonumber\\
&	\leq \frac{1}{m!\tau}\Bigg(
		\epsilon^{m}\int_{|y|\leq \epsilon} \td^{d} y \, p(x + y|x)
	\nonumber\\
&	+	\epsilon^{m+d}C\int_{|\xi|\geq 1} \td^{d} \xi \, \exp(-|\xi|) |\xi_{i_1}|\cdots |\xi_{i_m}|
	\Bigg)
	\nonumber\\
&	\leq \frac{\epsilon^{m}}{m!\tau}\left(1 +	C'\epsilon^{d}\right)
\end{align}
where $C'$ is a constant. This is sufficient to guarantee that the higher terms in $\epsilon$ vanish in the continuum limit, i.e. when $\epsilon \to 0$ with $\epsilon^{2}/\tau$ fixed.


%

\end{document}